# Electron-Phonon Scattering in Metallic Single-Walled Carbon Nanotubes

Ji-Yong Park, Sami Rosenblatt, Yuval Yaish, Vera Sazonova, Hande Üstünel, Stephan Braig,
T. A. Arias, Piet W. Brouwer, and Paul L. McEuen

*Laboratory of Atomic and Solid-State Physics, Cornell University, Ithaca, NY 14853*

Electron scattering rates in metallic single-walled carbon nanotubes are studied using an atomic force microscope as an electrical probe. From the scaling of the resistance of the same nanotube with length in the low and high bias regimes, the mean free paths for both regimes are inferred. The observed scattering rates are consistent with calculations for acoustic phonon scattering at low biases and zone boundary/optical phonon scattering at high biases.

Phonons in carbon nanotubes (NTs) have been an active area of research in recent years [1]. The effects of phonons have been measured in thermal transport [2-4], Raman scattering [1], and electrical transport [5-7]. Electron-phonon coupling in NTs is predicted to lead to a Peierls instability [8-11] and superconductivity [12,13] and contributes to the thermopower [14] and the resistivity [5,15-17] of NTs. At room temperature, the main origin of the resistivity at low bias in high quality metallic single-walled nanotubes (SWNTs) is believed to be scattering by acoustic phonons [5]. The scattering is weak, resulting in long mean free paths at room temperature. Both measurements and calculations put the mean free path in the range of a few hundred nm to several μm [5,6,15,17-19]. At high biases, electrons gain enough energy to emit optical or zone boundary phonons. Yao *et al.* showed that this scattering is very effective, leading to a saturation of the current ~20 μA [7] at high biases. This work indicated that the fundamental high-energy phonon scattering rate was rapid, but did not determine its value.

In this Letter, we study electron-phonon scattering in metallic SWNTs in both the low and high bias regime. We investigate the scaling of the resistance in a single NT for lengths ranging from 50 nm to 10 μm by using a tip of an atomic force microscope (AFM) as a movable electrode. For low biases, we find a length-independent resistance for $L < 200$ nm, indicating ballistic transport in the NT. At longer lengths, the scaling of the resistance with length is used to infer a low-bias mean free path of $\ell_{low} \approx 1.6$ μm, consistent with previous measurements and calculations. For high biases and long channel lengths, the current saturates as in the previous study [7]. However for short channel lengths, $L < 500$ nm, the current does not saturate, but rather grows linearly with increasing bias. From measurements at different lengths, we infer the high-bias mean free path, $\ell_{high} \approx 10$ nm, over 100 times shorter than $\ell_{low}$. We present calculations of the electron-phonon scattering rate due to acoustic phonons, relevant for low biases, and optic and zone boundary phonon emission, relevant at high biases. We obtain good agreement with our experimental results in both cases.

The SWNTs used in this study were grown by chemical vapor deposition at lithographically-defined catalyst sites on a degenerately-doped Si substrate with a 200 nm-thick oxide [20,21]. Metal electrodes were formed on the NT by evaporating 50 nm Au with 5 nm Cr adhesion layer or 30 nm Au without adhesion layer, followed by liftoff. Devices with direct Au contacts show better contact resistances after initial fabrication. Devices with Cr/Au contacts were annealed at ~600 °C after the fabrication to enhance the NT-metal contact. All the measurements reported here were performed on metallic SWNTs, as determined by the gate voltage dependence of the conductance and a saturation current of 19 – 25 μA at high biases and long channel lengths [22].

For this study, an AFM operating in ambient conditions with the capability of simultaneous transport measurements was used [23]. Figure 1 shows a schematic of the three-probe measurement geometry [24]. In addition to the two lithographic electrodes, an Au-coated AFM tip serves as a third electrode by making physical and electrical contact with the NT [25,26]. A bias voltage $V_{sd}$ is applied to the source electrode (the left electrode in Fig. 1), and the resulting current is measured by a current amplifier attached to the AFM tip, which serves as the drain electrode. This yields $I$ as a function of $V_{sd}$. We also use the second lithographic contact (right electrode in Fig. 1) as a voltage probe to measure the voltage drop $V_{tt}$ at the tip-tube drain junction.

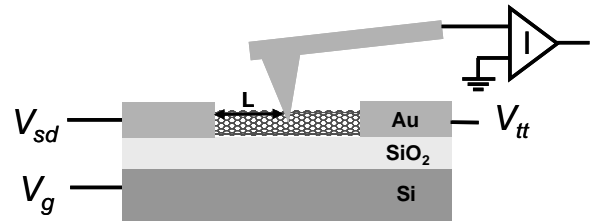

FIG. 1. A schematic of the three-probe measurement setup. The two electrodes and the Au-coated AFM tip serve as a voltage source (left electrode), voltage probe (right electrode) and current probe (AFM tip), respectively. The active length $L$ of the SWNT device can be changed by moving the AFM tip.

From this, we can deduce the voltage $V = V_{sd} - V_{tt}$ and hence the $I$-$V$ curve that corresponds to just the left contact and the SWNT segment between this contact and the tip. By moving the tip to different locations along the tube, $I$-$V$ measurements of the same SWNT for different channel lengths $L$ are obtained [27]. Since the left contact is always the same, and the voltage drop associated with the right AFM contact is subtracted, measurements at different channel lengths can be directly compared. The shortest channel length we can measure in this way depends on the stability of the AFM and the size of the tip, but is typically $L \approx 50$ nm.

Figure 2 is the result from measurements of $I$ vs. $V$ on a 10 µm-long metallic SWNT at six different tip-contact separations $L$. We first concentrate on the low bias region where the $I$-$V$ characteristic is linear. In Fig. 3(a), the low bias resistance $R_{low} = dV/dI$ is plotted for each length. For $L$ between 50 nm and 200 nm, $R_{low}$ is almost constant, but it increases linearly with $L$ for longer lengths. The slope of this line is the 1D resistivity $\rho = dR_{low}/dL \approx 4$ k$\Omega$/µm. Similar resistivity is found for several other metallic SWNTs.

The resistivity of a 1D channel with 4 subbands in the incoherent limit is given by [28]:

$$\rho = \left(h/4e^2\right)(1/\ell), \quad (1)$$

where $\ell$ is the electron mean free path for backscattering. From the measured slope, we infer $\ell_{low} \approx 1.6$ µm. For measurements with $L \ll \ell_{low}$, the transport is essentially ballistic. The measured resistance is just the contact resistance associated with the source contact and is constant.

Now we consider the high bias regime. The slope of the $I$-$V$ curve decreases with increasing $V$. For long channel lengths, the current saturates to an approximately constant value of ~20 µA as reported previously [7]. For shorter lengths $L <$ 500 nm, however, different behavior is observed. The current again increases linearly with $V$ at high bias, but with a slope much lower than at low bias.

The inverse slopes of $I$-$V$ curves at high bias, $R_{high} = dV/dI$, from 4 devices with diameters in the range 1.8nm $< d <$ 2.5nm are shown in Fig 3(b). $R_{high}$ scales linearly with $L$, with a slope $dR_{high}/dL = 800$ k$\Omega$/µm. This high-bias resistivity is 200 times larger than at low bias. If we use Eq. (1) to infer a mean free path, we obtain $\ell_{high} \approx 10$ nm.

We now compare these results to expectations for electron-phonon scattering in NTs. The electron-phonon coupling is described by the Hamiltonian

$$H_{ep} = \sum_{\mathbf{k},\mathbf{q}} D^{\alpha}_{\mathbf{k},\mathbf{q}} c^{\dagger}_{\mathbf{k}+\mathbf{q}} c_{\mathbf{k}} u^{\alpha}_{\mathbf{q}}, \quad (2)$$

$$u^{\alpha}_{\mathbf{q}} = \sqrt{\frac{\hbar}{2L\rho\Omega^{\alpha}_{\mathbf{q}}}} \left(b^{\alpha}_{\mathbf{q}} + b^{\alpha\dagger}_{-\mathbf{q}}\right),$$

where $\mathbf{q}$ and $\alpha$ label the phonon wavevector and branch, respectively, $u$ is the phonon displacement, $c^{\dagger}$ and $c$ ($b^{\dagger}$ and $b$) are electron (phonon) creation and annihilation operators, respectively, $L$ is the length of a NT, and $\rho$ is the mass density. For application to electron transport, the electron wavevectors $\mathbf{k}$ and $\mathbf{k} + \mathbf{q}$ are taken in the vicinity of the Fermi points of the NT. Electron-phonon scattering of electrons from an initial state with wavevector $\mathbf{k}$ and energy $E_{\mathbf{k}}$ to a final state with wavevector $\mathbf{k} + \mathbf{q}$ and energy $E_{\mathbf{k}+\mathbf{q}}$ must conserve energy,

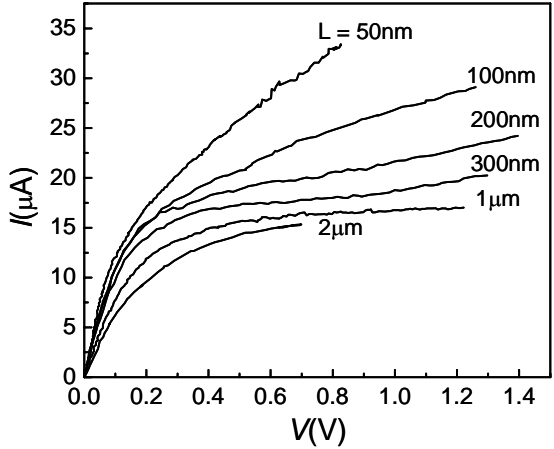

FIG. 2. The current $I$ versus voltage $V$ at different lengths for a SWNT device of overall length 10 µm and tube diameter 1.8 nm. The length $L$ of the tube measured is indicated next to each curve.

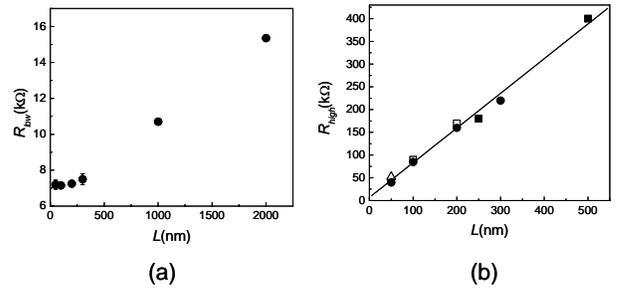

Figure 3. (a) The low bias resistance $R_{low} = dV/dI$ near zero bias as a function of $L$ for the SWNT in Fig. 2. (b) The high bias resistance $R_{high} = dV/dI$ versus $L$, from four different SWNTs. The diameters of the SWNTs are 1.8 nm (filled circles, from the device in Fig. 2), 2 nm (filled square, empty square), and 2.5 nm (empty circle). The line is a linear fit to the data.

$$E_{\mathbf{k}+\mathbf{q}} = E_{\mathbf{k}} \pm \hbar\Omega_{\mathbf{q}}^{\alpha}, \quad (3)$$

where the plus sign refers to absorption of a phonon with wavevector **q** and the minus sign to emission of a phonon with momentum –**q**. Examples of scattering processes that meet these criteria in NT are shown schematically in Fig. 4.

For acoustic phonons, one has $D_{\mathbf{k}_F,\mathbf{q}}^{\alpha} = \Xi^{\alpha}q$, where $\Xi^{\alpha}$ is the deformation potential. Using Fermi's golden rule along with a thermal occupation of phonon states ($\hbar\Omega_{\mathbf{q}}^{\alpha} \ll kT$ for acoustic phonons) and combining contributions from emission and absorption of acoustic phonons, one arrives at the electron-acoustic phonon scattering rate

$$\frac{1}{\tau_{ac}} = 2\frac{2\pi}{\hbar}\Xi^2\left(\frac{k_B T}{2\rho v_s^2}\right)\frac{1}{hv_F}, \quad (4)$$

where $v_s = \Omega_{\mathbf{q}}/q$ is the acoustic phonon velocity. The maximum deformation potential for an acoustic phonon from band gap change in metallic SWNTs with uniaxial or torsional strain is $\Xi \approx 5$ eV [29]. Then from Eq. (4), we compute the scattering time, $\tau_{ac} \approx 3.0\times10^{-12}$ s for a SWNT with diameter 1.8 nm and the corresponding mean free path, $\ell_{ac} = v_F\tau_{ac} \approx 2.4$ μm. This calculated acoustic phonon scattering rate is comparable to previous calculations [5,15].

Since there are two possible origins of electron scattering at low bias, disorder and acoustic phonons, the experimentally measured value $\ell_{low}$ is a lower limit for the acoustic phonon scattering mean free path: $\ell_{low} < \ell_{ac}$. The data is therefore consistent with the theory. These measurements confirm the weak electron-acoustic phonon scattering and long mean free path values previously reported in SWNTs [5,6,18,19].

We now turn to the case of optical and zone boundary phonons, where the scattering length has not been determined previously.

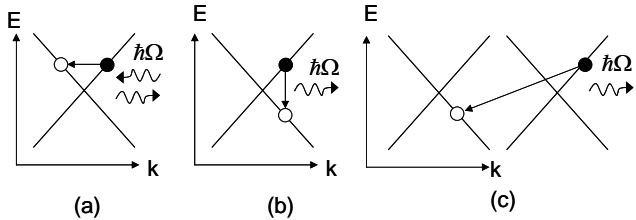

Figure 4. Diagrams showing electron-phonon scattering in a metallic SWNT by phonons with energy $\hbar\Omega$ that satisfies energy and momentum conservation. (a) Emission and absorption of acoustic phonon with low energy and small momentum. (b) Emission of optical phonon (high energy and small momentum) and (c) zone-boundary phonon (high energy and large momentum).

Since no such phonons are present at room temperature ($\hbar\Omega \gg k_B T$), scattering from optical or zone boundary phonons will involve emission only if we assume that any emitted phonons relax instantaneously. These processes are shown in Figs. 4(b) and (c). According to Fermi's golden rule, the scattering rate from spontaneous phonon emission is

$$\frac{1}{\tau_{\mathbf{q}}^{\alpha}} = \frac{2\pi}{\hbar}\left|D_{\mathbf{k}_F,\mathbf{q}}^{\alpha}\right|^2\left(\frac{\hbar}{2\rho\Omega_{\mathbf{q}}^{\alpha}}\right)\frac{1}{hv_F}. \quad (5)$$

Using a tight binding model [30], we calculated the relevant matrix elements $D_{\mathbf{k},\mathbf{q}}^{\alpha}$ in the electron-phonon Hamiltonian by considering the creation of a band gap in NT with respect to the atomic displacements corresponding to the phonon mode. For optical phonons (**q** = 0), we found that only displacements compatible with the longitudinal mode lead to a band gap [31]. From this model, we obtain $D \approx 12.8$ eV/Å, corresponding to the scattering time $\tau_{op} \approx 2.3\times10^{-13}$ s and the mean free path $\ell_{op} \approx 180$ nm. For the zone boundary phonon (**q** = -2**k**$_F$), we found that only one optical mode gives a significant band gap while contributions of other optical and acoustic modes are small. For this mode, we found $D \approx 25.6$ eV/Å, yielding $\tau_{zo} \approx 4.6\times10^{-14}$ s and $\ell_{zo} \approx 37$ nm. Unlike the acoustic scattering mean free path, which depends on the type of nanotube, the mean free paths $\ell_{op}$ and $\ell_{zo}$ do not depend on the type of NT, as long as the NT is metallic.

In principle, to calculate the effect of high energy scattering processes on the *I-V* characteristics, the Boltzmann equation must be solved self-consistently to determine the occupancies of the initial and final states. In the limiting case of high bias, however, simple arguments can be given, which are sufficient for interpretation of the data.

Following Ref. 5, an electron must first accelerate a length $\ell_T$ in the electric field $\varepsilon$ to attain sufficient excess energy to emit an optical phonon ($\hbar\Omega \approx 0.2$ eV) or zone boundary phonon ($\hbar\Omega \approx 0.16$ eV). This length is given by $\ell_T = \hbar\Omega/e\varepsilon = (\hbar\Omega/eV)\cdot L$. Once it attains this energy, it will scatter on a length scale of high energy phonon scattering mean free path ($\ell_{hp}$). The total length the electron traverses before scattering is, therefore, of order $\ell_T + \ell_{hp}$, as illustrated in Fig 5.

For large enough $L$, $\ell_T \gg \ell_{hp}$, the scattering length will be determined by $\ell_T$. The high bias resistance is then $R = (h/4e^2)(L/\ell_T) + R_{low} = V/I_0 + R_{low}$, which implies a current saturation with $I_o = (4e/h)\hbar\Omega$ [7]. This is valid for the longer channel lengths in Fig. 2.

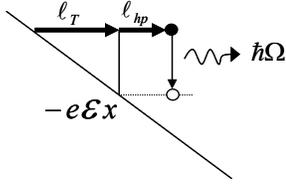

Figure 5. A schematic showing the two relevant length scales for high bias electron-phonon scattering. The total scattering length is the sum of $\ell_T$, the distance electron must accelerate in the electric field $\varepsilon$ to reach a threshold energy $\hbar\Omega$, and $\ell_{hp}$, the high energy phonon scattering mean free path.

On the other hand, if $L$ is sufficiently small, $\ell_T \ll \ell_{hp}$, the scattering length will be determined by $\ell_{hp}$. Then, the high bias resistance is $R_{high} = (h/4e^2)(L/\ell_{hp}) + R_{low}$. In short channels, then, the measured scattering length should be $\ell_{hp}$ [32]. Since both optical and zone boundary phonons contribute, we can deduce the scattering length from our calculations: $\ell_{op}^{-1} + \ell_{zo}^{-1} \approx 30$ nm.

This theoretical value is in reasonable agreement with the experimental finding: $\ell_{high} \approx 10$ nm. The difference could arise from uncertainties in the parameters of the theory or from additional scattering mechanisms not considered here. These include multiple phonon scattering or stimulated phonon emission from high energy phonons created by other scattering events. The latter process will be important if the phonons can not rapidly escape the NT into the substrate. Measurements at high bias on suspended NTs [33] should shed light on this issue.

In conclusion, we have investigated the scaling with channel length of the *I-V* characteristics of metallic SWNTs. The mean free paths for electron-acoustic phonon scattering ($\ell_{ac} > 1.6$ μm) and high energy phonon scattering ($\ell_{high} \approx 10$ nm) at room temperature were measured and compared to theory. In addition to their fundamental interest, these results are important for uses of SWNTs in high current applications such as interconnects or transistors operating in the saturation region. Our results show that the scattering length at high biases is more than 100 times shorter than at low bias. Nevertheless, in short devices, the current carried by a SWNT can significantly exceed the previously reported limit of 25 μA [7].

This work was supported by the NSF Center for Nanoscale Systems, the Packard Foundation, the MARCO/DARPA FRC-MSD, which is funded at MIT under contract 2001-MT-887 and MDA972-01-1-0035, and the Cornell Center for Materials Research under NSF Grant No. DMR0079992. Sample fabrication was performed at the Cornell node of the National Nanofabrication users Network, funded by NSF.